\documentclass[aps,a4paper, preprint, superscriptaddress,preprintnumbers,floatfix,nofootinbib,amsmath,amssymb]{revtex4-1}

\usepackage{url}
\usepackage{hyperref}
\usepackage{color}
\usepackage{cancel}
\usepackage{cleveref}
\usepackage{subfig}
\usepackage{soul}
\usepackage[normalem]{ulem}

\usepackage{amstext,amssymb}
\usepackage{amsmath}
\usepackage{graphicx}
\usepackage{xspace}
\usepackage{color}
\usepackage{units}
\usepackage[T1]{fontenc}
\usepackage{amsmath,bm}
\usepackage{nicefrac}
\usepackage{slashed} 
\usepackage{multirow}
\newcommand{\be}{\begin{equation}}
\newcommand{\ee}{\end{equation}}
\newcommand{\bea}{\begin{eqnarray}}
\newcommand{\eea}{\end{eqnarray}}

\newcommand{\mycomment}[1]{}

\begin{document}

\title{Impact of different neutrino decoherence formalisms at the future long-baseline experiments}

\author{Rudra Majhi}
\email{rudra.majhi95@gmail.com}
\affiliation{Nabarangpur College, Nabarangpur - 764059, Odisha, India}  

\author{Koushik Pal}
\email{palkoushik01@gmail.com}
\affiliation{School of Physics,  University of Hyderabad, Hyderabad - 500046,  India}     
\author{Monojit Ghosh}
\email{mghosh@irb.hr}
\affiliation{Center of Excellence for Advanced Materials and Sensing Devices, Ru{\dj}er Bo\v{s}kovi\'c Institute, 10000 Zagreb, Croatia}     

\author{Rukmani Mohanta}
\email{rmsp@uohyd.ac.in}
\affiliation{School of Physics,  University of Hyderabad, Hyderabad - 500046,  India}

\begin{abstract}

In this paper, we have studied the impact of two different formalisms of quantum decoherence in determining the sensitivities of the two future long-baseline experiments DUNE and P2SO. In formalism-A, we will assume that the decoherence matrix is defined in a matter mass eigenstate basis which is the basis that diagonalizes the  Hamiltonian for neutrinos in matter, with a constant matter density. In formalism-B, we will define the decoherence matrix in the vacuum mass eigenstate basis and then rotate  it to matter mass basis via an unitary transformation. By using different values of the decoherence parameter $\Gamma$, we will show how these two formalisms differ at the probability level and then we will demonstrate how the sensitivities can differ at the $\chi^2$ level. Our results show that if the values of $\Gamma$ are small, then these two formalisms yield same probability in vacuum. However, if the values of $\Gamma$ is large or if there is strong matter effect, then these two formalisms yield very different results.

\end{abstract}

\maketitle

\section{Introduction}
Experimental observations of neutrino oscillations in several dedicated experiments have firmly established that neutrinos possess nonzero masses and that oscillations arise from quantum interference among different neutrino mass eigenstates. In the present landscape of neutrino physics, one of the major goals is to determine the leptonic CP-violating phase, as well as to resolve the issues related to the octant of $\theta_{23}$ and the ordering of neutrino masses~\cite{Capozzi:2018ubv}.

The standard  oscillation paradigm assumes completely coherent propagation of neutrino mass states. However, subleading effects beyond the standard scenario may alter oscillation patterns. One such effect is \textit{decoherence}, namely the loss or degradation of quantum coherence between different components of the neutrino state induced either by interactions with an external environment or by intrinsic stochastic fluctuations. Decoherence acts like a damping factor in the interference between neutrino mass eigenstates, typically through exponential suppression factors that alter the oscillatory behavior and consequently modify the flavor transition probabilities.
Phenomenologically, decoherence is introduced by an exponential damping factor, $e^{-\Gamma L}$, where $\Gamma$ denotes the decoherence strength and $L$ is the baseline. The presence of decoherence can also shift or distort the effective oscillation frequencies \cite{Benatti:2000ph, Fogli:2007tx, Ohlsson:2000mj,Blennow:2005yk,Blennow:2005qj}.

The decoherence parameter $\Gamma$ may depend on the neutrino energy and is often parametrized in the power-law form as $\Gamma \propto E_\nu^{\,n}$, where $n$ is an integer~\cite{Lisi:2000zt}. In this work, we consider the case $n=0$, corresponding to an energy-independent decoherence scale. Possible origin of such effects have been discussed in various contexts, including quantum gravity or space-time foam~\cite{Giddings:1988cx,Amelino-Camelia:1996bln,Garay:1999cy,Ellis:1983jz, Domi:2024ypm}, stochastic matter effects~\cite{Akhmedov:2010ms,Fogli:2007tx, Stuttard:2020qfv}, wave-packet separation~\cite{Akhmedov:2022bjs}, environmental interactions~\cite{Benatti:2000ph,Lisi:2000zt}, and nonunitary evolution described by open quantum systems~\cite{Lindblad:1975ef}.
Quantum gravity suggests that space-time may not be smooth at the Planck scale, but instead exhibits stochastic fluctuations (often referred to as space-time foam). These fluctuations can couple to quantum fields and, at energies below the Planck scale, induce corrections to the quantum evolution of particles. As a consequence, the system may no longer evolve unitarily, effectively violating standard quantum mechanical probability conservation \cite{Wheeler:1955zz, Giddings:1988cx}. 

Neutrinos, being light  particles with extremely small mass, provide an ideal platform to probe tiny decoherence effects with inordinate sensitivity. Loss of coherence does not forbid flavor oscillations; instead, it modifies the oscillation pattern through its dependence on the neutrino energy and the propagation distance. Therefore, experiments in which neutrinos travel long distances are particularly suitable for studying decoherence effects. In this context, several studies have been performed using solar neutrinos, atmospheric neutrinos \cite{KM3NeT:2024jji, DeRomeri:2023dht, Fogli:2007tx, Coloma:2018idr, Gonzalez-Garcia:2005ryx, ICECUBE:2023gdv, Acharya:2025rnw}, long-baseline accelerator neutrinos \cite{DeRomeri:2023dht, Gomes:2020muc, Barenboim:2024wdn, ESSnuSB:2024yji, Bera:2024hhr, Carrasco:2018sca, BalieiroGomes:2018gtd,Coelho:2017zes, Oliveira:2016asf, Coelho:2017byq}, reactor neutrinos \cite{DeRomeri:2023dht, deGouvea:2020hfl, deGouvea:2021uvg, Krueger:2023skk, Akhmedov:2022bjs}, and supernova neutrinos \cite{dosSantos:2023skk, Ternes:2025mys}.

In recent times there has been some discussion about how to compute the neutrino oscillation probabilities in presence of decoherence in matter. One of the popular ways is to assume that the decoherence matrix can be defined in a matter mass eigenstate basis as diagonal and energy independent \cite{BalieiroGomes:2018gtd} (we call this as  `formalism-A'.). The matter mass eigenstates are the basis that diagonalize the  Hamiltonian for neutrinos in matter, with a constant matter density. However, it was argued that these assumptions are fulfilled only in a few cases \cite{Carpio:2017nui}. The correct way is to define the decoherence matrix in the vacuum mass eigenstate basis and one requires to rotate it to matter mass basis via an unitary transformation (we call this `formalism-B'). Further, they showed that in formalism A, there appears a purported peak in the appearance channel probability. In reply to this, it was commented that under certain conditions, decoherence can be regarded as arising from the same matrices in both the contexts, such that it preserves the physical interpretation wherein the decoherence effect acts only on the quantum interference terms \cite{BalieiroGomes:2018gtd}. In this paper, our goal is to consider the neutrino oscillation probabilities computed using both the formalisms for a given decoherence matrix and study their implications in the two future long-baseline experiments—DUNE and Protvino-to-Super-ORCA (P2SO), wherein the matter effect is quite substantial. First, we will show at the probability level, under what condition the two formalisms agree and when they tend to disagree. Further, we will compute the bounds on the decoherence parameter and its impact on the mass ordering, octant of $\theta_{23}$ and CP violation using both the formalisms and quantify the differences. Finally, we will also comment about the effect of the purported probability peak which appears in Formalism A, in estimating the sensitivity of the above mentioned experiments. Our study will help to understand the importance of choosing a particular formalism when estimating the sensitivity of quantum decoherence in neutrino oscillation experiments with strong matter effect. 

The paper is organized as follows. In Sec.~\ref{theoretical}, we layout the differences between the above-mentioned two formalisms. Section~\ref{simulation} describes the simulation methodology and the experimental configurations employed in this study. The results and discussion are presented in Sec.~\ref{result}. Finally, we summarize our findings in Sec.~\ref{summary}.

\section{The two formalisms for computing neutrino oscillation probabilities in presence of quantum decoherence}
\label{theoretical}
In an open quantum system approach for neutrino propagation, in which the neutrino states may interact weakly with an unknown environment. In this framework, the evolution of the neutrino system acquires damping terms that attenuate the oscillation amplitudes. For a pure state, the evolution is governed by the Schrödinger equation, whereas for mixed states one employs the Liouville-von Neumann equation
\begin{equation}
    \frac{d\rho^{m}(t)}{dt} = -\,i\left[\,\mathcal{H},\,\rho^{m}(t)\,\right],
\end{equation}
where $\mathcal{H}$ denotes the Hamiltonian of the system and $\rho^{m}$ is the density matrix on the mass basis, defined as
\begin{equation}
    \rho^{m} = \sum_{j} p_j\, |\nu_j \rangle \langle \nu_j |\;,
\end{equation}
$p_j$ being the probability  that the neutrino  is in the mass state $|\nu_j \rangle$.
In the presence of decoherence, the dynamics becomes nonunitary and can be consistently described by the Lindblad master equation \cite{Gorini:1975nb, Lindblad:1975ef}
\begin{equation}
    \frac{d\rho^{m}(t)}{dt} = -\,i\left[{\cal H}_{\text{eff}}^{m},\,\rho^{m}(t)\right] + 
    \mathcal{D}\left[\rho^{m}(t)\right],
    \label{eq:LME}
\end{equation}
where $\mathcal{D}[\rho]$ encodes the dissipative interaction between the neutrino system and the environment. The dissipator can be parametrized as
\begin{equation}
    \mathcal{D}[\rho^{m}] = \frac{1}{2}\sum_{n=1}^{N^2-1}
    \Big( \left[V_n ,\, \rho^{m} V_n^\dagger\right] +
          \left[V_n \rho^{m},\, V_n^\dagger\right] \Big)\;,
    \label{eq:D-term}
\end{equation}
with $\{V_n\}$ a set of Lindblad operators for $N$ neutrino generations. For three neutrino generations, $n$ varies from $1,2,...,8$. The presence of these operators renders the evolution nonunitary. To ensure complete positivity and a monotonically increasing von Neumann entropy, one typically requires the $V_n$ to be Hermitian. The solution to the Lindblad master equation is 
\begin{equation}
    \rho^{m}(t) = e^{(M_H + M_D)t}\rho^{m}(0).
\end{equation}
For three flavor case, $\rho^{m}$ is an eight-dimensional column vector, $M_H$ and $M_D$ involve the Hamiltonian and all the decoherence parameters and are of $8\times 8$ matrices. The probability of $\nu_\alpha \to \nu_\beta $  can be expressed as 
\begin{eqnarray}
    P(\nu_\alpha \to \nu_\beta ) = \frac{1}{N}+ \frac{1}{2}(\rho^{m}_\beta)^T \rho^{m}_\alpha(t).
    \label{prob-exp}
\end{eqnarray}
For the case of 3 flavors,  $N=3$ and for ultrarelativistic neutrinos one can  express $t\simeq L$, with $L$ being the baseline. Let us now focus on the dissipative matrix $\mathcal{D}$. In three flavor case,  the matrix can be written in terms of the Gell-Mann matrices $\lambda_j$ as
\begin{equation}
    \mathcal{D} = \mathcal{D}_{jk}\rho^m_k \lambda_j,
\end{equation}
where the $\rho^m_k$ are the elements of the density matrix $\rho^m$ in three flavor case. 

There  are several ways of implementation of the formalism for the decoherence. In formalism-A, one takes the diagonal dissipator form in the matter mass basis as \cite{Oliveira:2013nua, BalieiroGomes:2016ykp, Oliveira:2016asf, BalieiroGomes:2018gtd, DeRomeri:2023dht}
\begin{equation}
     \mathcal{D}_{jk} = - \mathrm{diag}(\Gamma_{21},\Gamma_{21},0,\Gamma_{31},\Gamma_{31},\Gamma_{32},\Gamma_{32},0),
\end{equation}
where the $\Gamma_{ij}$ are the decoherence parameters with~\cite{ESSnuSB:2024yji}
\begin{equation}
\Gamma_{31} = \Gamma_{21} + \Gamma_{32} - 2 \sqrt{\Gamma_{21} \Gamma_{32}}.
\label{rln}
\end{equation}
Here  ${\cal D}_{33}=-\Gamma_{33}=0$, and ${\cal D}_{88}=- \Gamma_{88}=0$ are known as the relaxation terms.  In this formalism, the decoherence parameters appear as damping factors of the form $\exp(-\Gamma_{ij} L)$ in the oscillation probabilities, with $L$ being the baseline length. 

Under the assumption of ultrarelativistic neutrinos ($t \simeq L$), the oscillation probability can be derived from Eq.~\ref{prob-exp} as \cite{Gago:2002na, Coloma:2018idr}, 
\begin{equation}
\begin{aligned}
        P_{\alpha \beta}(L) &= \delta_{\alpha \beta} - 2\sum_{j > k} Re \left( \tilde{U}_{\beta j} \tilde{U}_{\alpha j}^* \tilde{U}_{\alpha k} \tilde{U}_{\beta k}^* \right) + 2\sum_{j > k} Re \left( \tilde{U}_{\beta j} \tilde{U}_{\alpha j}^* \tilde{U}_{\alpha k} \tilde{U}_{\beta k}^* \right) \exp(-\Gamma_{jk} L) \cos \left (\frac{\Delta \tilde{m}_{jk}^2}{2E}L \right ) \\& + 2\sum_{j > k} Im \left( \tilde{U}_{\beta j} \tilde{U}_{\alpha j}^* \tilde{U}_{\alpha k} \tilde{U}_{\beta k}^* \right) \exp(-\Gamma_{jk} L) \sin\left (\frac{\Delta \tilde{m}_{jk}^2}{2E}L \right ),
        \end{aligned}
        \label{eq:Pab}
\end{equation} 
where $ \tilde{U}_{i j}$ and $\Delta \tilde{m}_{jk}$ denote the effective PMNS matrix elements and effective mass-squared differences, respectively, describing neutrino propagation in matter.

In formalism-B~\cite{Carpio:2017nui}, the decoherence matrix $\mathcal{D}$ is, in general, defined in diagonal form in the vacuum mass-eigenstate basis. When neutrino propagation in matter is considered, an appropriate rotation of the decoherence matrix $\mathcal{D}$ into the effective matter basis is required. 

References.~\cite{Carpio:2017nui,Carpio:2018gum} show that, for the decoherence matrix in the vacuum basis:

\begin{align}
\mathcal{D}_{jk} &= -\mathrm{diag}\!\left(
\Gamma_2,\, \Gamma_2,\, 0,\,
\Gamma_4,\, \Gamma_4,\,
\Gamma_4,\, \Gamma_4,\, 0
\right),
\label{form-B}
\end{align}
 where $\Gamma_2$ and $\Gamma_4$ are the decoherence parameters.
Starting from the Lindblad master equation and changing $M_D$ from vacuum mass basis to matter mass basis via a unitary transformation, one can derive the neutrino oscillation probability formulas as: 
\begin{eqnarray}
P_{\nu_\mu \to \nu_e} 
&=& P^{(0)}_{\nu_\mu \to \nu_e} 
+ \frac{\bar{\Gamma}_2}{2} \cos^2\theta_{23} \sin^2 2\theta_{12}
\left[1-\frac{\bar{\Gamma}_2}{2}\sin^2 2\theta_{12} -
\frac{\sin^2(A\Delta)}{2A^2 \Delta^2} \cos^2 2\theta_{12}
\right] \nonumber\\
&
+& \frac{\bar{\Gamma}_2 \theta_{13} \sin 2\theta_{23} \sin 4\theta_{12}}
{4(1-A)A\Delta}
\left[
\sin(A\Delta)\cos(\delta_{\rm CP} + A\Delta)
- A^2 \sin\Delta \cos(\delta_{\rm CP} + \Delta)
\right] \nonumber\\
&-& \frac{\alpha \bar{\Gamma}_2}{2A^2 \Delta}
\cos 2\theta_{12} \cos^2\theta_{23} \sin^2 2\theta_{12}
\left(
\sin (2A\Delta) - 2A\Delta
\right),
\label{pmue-for-B}
\end{eqnarray}
\begin{eqnarray}
P_{\nu_\mu \to \nu_\mu}
&=& P^{(0)}_{\nu_\mu \to \nu_\mu}
- \frac{\bar{\Gamma}_4}{2} \sin^2 2\theta_{23}\cos(2\Delta)\left(1- \frac{\bar{\Gamma}_4}{2} \right)
 \nonumber\\
&+& \frac{\bar{\Gamma}_2^2}{4}\cos^4\theta_{23}
\left[
\frac{\sin^2(A\Delta)}{4A^2 \Delta^2}\sin^2 4\theta_{12}
+  \sin^4 2\theta_{12}
\right] \nonumber\\
&+& \frac{\bar{\Gamma}_2  \alpha \cos 2\theta_{12} \cos^4\theta_{23} \sin^2 2\theta_{12}}
{2A^2 \Delta}
\left(
\sin 2A\Delta - 2A\Delta
\right)\nonumber\\
&
-& \frac{\bar{\Gamma}_2 \theta_{13} \sin 2\theta_{23} \cos^2\theta_{23} \sin 4\theta_{12} \cos\delta_{\rm CP}}
{4(1-A)\Delta}
\left(
\sin 2A\Delta - A^2 \sin 2\Delta
\right),
\end{eqnarray}
where $ P^{(0)}_{\nu_\mu \to \nu_\mu}$ and $ P^{(0)}_{\nu_\mu \to \nu_e}$ are the survival and appearance probabilities in standard scenario without decoherence parameter ($\Gamma = 0$), $\bar{\Gamma}_i = \Gamma_i L$, 
$\Delta = \dfrac{\Delta m^2_{31} L}{4E}$, 
 $A = \dfrac{2\sqrt{2}G_F n_e E}{\Delta m^2_{31}}$, and $\alpha = \Delta m^2_{21}/\Delta m^2_{31}$. Here $G_F$, $n_e$ and $E$ are the Fermi coupling constant, electron number density and neutrino energy. The oscillation probabilities for antineutrinos can be obtained by performing the replacements $A \rightarrow -A$ and $\delta_{\rm CP} \rightarrow -\delta_{\rm CP}$.

At this stage, in order to make a comparison between these two formalisms, if we put $\Gamma_{31} = \Gamma_{32} = \Gamma_4$ and $\Gamma_{21} = \Gamma_2$ in Eq.(~\ref{rln}), we obtain $\Gamma_2 = 4\Gamma_4$. After redefining $\Gamma_2 = \Gamma$, we obtain a single form of $\mathcal{D}_{jk}$:
\begin{align}
\mathcal{D}_{jk} &= -\mathrm{diag}\!\left(
\Gamma,\, \Gamma,\, 0,\,
\frac{\Gamma}{4},\, \frac{\Gamma}{4},\,
\frac{\Gamma}{4},\, \frac{\Gamma}{4},\, 0
\right),
\quad
\quad.
\end{align}
for which both the probability formulas derived using two different formalisms can be applied.

\section{Simulation Details}
\label{simulation}

In this section, we describe the experimental configurations and numerical tools employed in our analysis. We focus on two future long-baseline accelerator experiments: P2SO and DUNE. The relevant setup details are summarized below.

\subsection{P2SO}
The Protvino-to-Super-ORCA (P2SO) proposal is one of the next-generation long-baseline experiments. In this setup, a neutrino beam will be produced at the U-70 synchrotron in Protvino, Russia, and directed toward the Super-ORCA detector located in the Mediterranean Sea, approximately $40\,\mathrm{km}$ off the coast of Toulon, France, with a baseline of $2595\,\mathrm{km}$. Further details of the experimental concept may be found in Refs.~\cite{Akindinov:2019flp,Singha:2022btw,Majhi:2022fed}.

For P2SO, a beam power of $450\,\mathrm{kW}$ is assumed, corresponding to an annual exposure of $4\times10^{20}$ protons on target (PoT). The Super-ORCA detector utilizes a medium with a density an order of magnitude larger than that of ORCA. The neutrino energy range relevant for physics studies extends from $0.2$ to $10\,\mathrm{GeV}$. In this work, we assume a total exposure of six years, symmetrically distributed between neutrino and antineutrino running.

\subsection{DUNE}

DUNE (Deep Underground Neutrino Experiment) is a forthcoming long-baseline neutrino oscillation facility to be operated at Fermilab in the United States. The detector complex comprises a near detector at Fermilab and a $40\,\mathrm{kt}$ liquid-argon time projection chamber (LArTPC) far detector, located in the Sanford Underground Research Facility in South Dakota. The experiment is designed to utilize a high-power beam, reaching $1.2\,\mathrm{MW}$ during its nominal phase. For simulation purposes, we employ the official GLoBES configuration files corresponding to the Technical Design Report (TDR)~\cite{DUNE:2021cuw}.

In our numerical study, we assume a total run time of 13 years, equally split between neutrino and antineutrino modes ($6.5+6.5$) years, yielding an annual exposure of $1.1\times10^{21}$ PoT. This corresponds to the staged operational scenario described in Ref.~\cite{DUNE:2020jqi}.

To simulate the event spectra, we use the GLoBES software package~\cite{Huber:2004ka,Huber:2007ji}. The sensitivity estimates are obtained via a Poissonian $\chi^2$ statistics,
\begin{equation}
\chi^2_{\rm stat}
= 2\sum_{i=1}^n \left[
N^{\rm test}_i - N^{\rm true}_i
- N^{\rm true}_i\ln\left(
\frac{N^{\rm test}_i}{N^{\rm true}_i}
\right)
\right],
\end{equation}
where $N^{\rm true}_i$ and $N^{\rm test}_i$ denote the expected and test event numbers in the $i{\rm th}$ energy bin, and $n$ is the total number of bins. Systematic uncertainties are incorporated using the pull method~\cite{Fogli:2002pt,Huber:2002mx}. The signal and background normalization uncertainties, as well as shape uncertainties (where applicable), assumed for P2SO and DUNE are given in Table~\ref{table_sys}. Shape systematics are only applied to P2SO in our setup.

The oscillation parameters are taken from the NuFIT\,6.1 global fit~\cite{Esteban:2024eli} and listed in Table~\ref{tab:osc-parameters}. In the numerical minimization, we vary the atmospheric mixing parameters $\theta_{23}$, $\Delta m^2_{31}$, and the CP phase $\delta_{\rm CP}$.

\begin{table} 
\centering
\begin{tabular}{|c|c|c|} \hline
Systematics     & P2SO          & DUNE  \\ \hline
Sg-norm $\nu_{e}$   & 5$\%$   & 2$\%$      \\ 
Sg-norm $\nu_{\mu}$    & 5$\%$            & 5$\%$ \\ 
Bg-norm    & 12$\%$     & 5$\%$ to 20$\%$\\ 
Sg-shape      & 11$\%$     & -\\ 
Bg-shape     & 4\% to 11$\%$       & - \\ 
\hline
\end{tabular}
\caption{Systematic errors and their values considered in our analysis. We have mentioned normalization error as ``norm$"$,  signal as ``Sg$"$ and background as ``Bg$"$.}
\label{table_sys}
\end{table}

\begin{table}[htbp!]
    \centering
    \begin{tabular}{|c|c|c|} \hline 
         Parameters&  Best fit value $\pm 1\sigma$& $3\sigma$\\ \hline 
         $\sin^2\theta_{12}$&  $0.3088^{+0.0067}_{-0.0066}$& $0.2893 \to 0.329$\\ \hline 
         $\sin^2\theta_{13}$&  $0.02248^{+0.00055}_{-0.00059}$& $0.02064 \to 0.02418$\\ \hline 
         $\sin^2\theta_{23}$&  $0.47^{+0.017}_{-0.014}$& $0.435 \to 0.584$\\ \hline 
         $\delta_{\rm CP} / ^\circ$&  $212^{+26}_{-36}$& $125 \to 365$\\ \hline 
         $\Delta m_{21}^2/ 10^{-5}~{\rm eV}^2$&  $7.537^{+0.094}_{-0.1}$& $7.236 \to 7.823$\\ \hline 
         $\Delta m_{31}^2/ 10^{-3}~{\rm eV}^2$&  $+2.511^{+0.021}_{-0.020}$& $+2.450 \to +2.576$\\ \hline
    \end{tabular}
\caption{Oscillation parameter values along with their $1\sigma$ and $3\sigma$ ranges, as reported in NuFIT~6.1~\cite{Esteban:2024eli}, for normal ordering.}
    \label{tab:osc-parameters}
\end{table}

\section{Result}
\label{result}
In this section we will present our result for the two formalisms of decoherence at probability level and at $\chi^2$ level, for the long-baseline experiments.

\subsection{Analysis at probability level}
\begin{figure}
\includegraphics[width=1\linewidth]{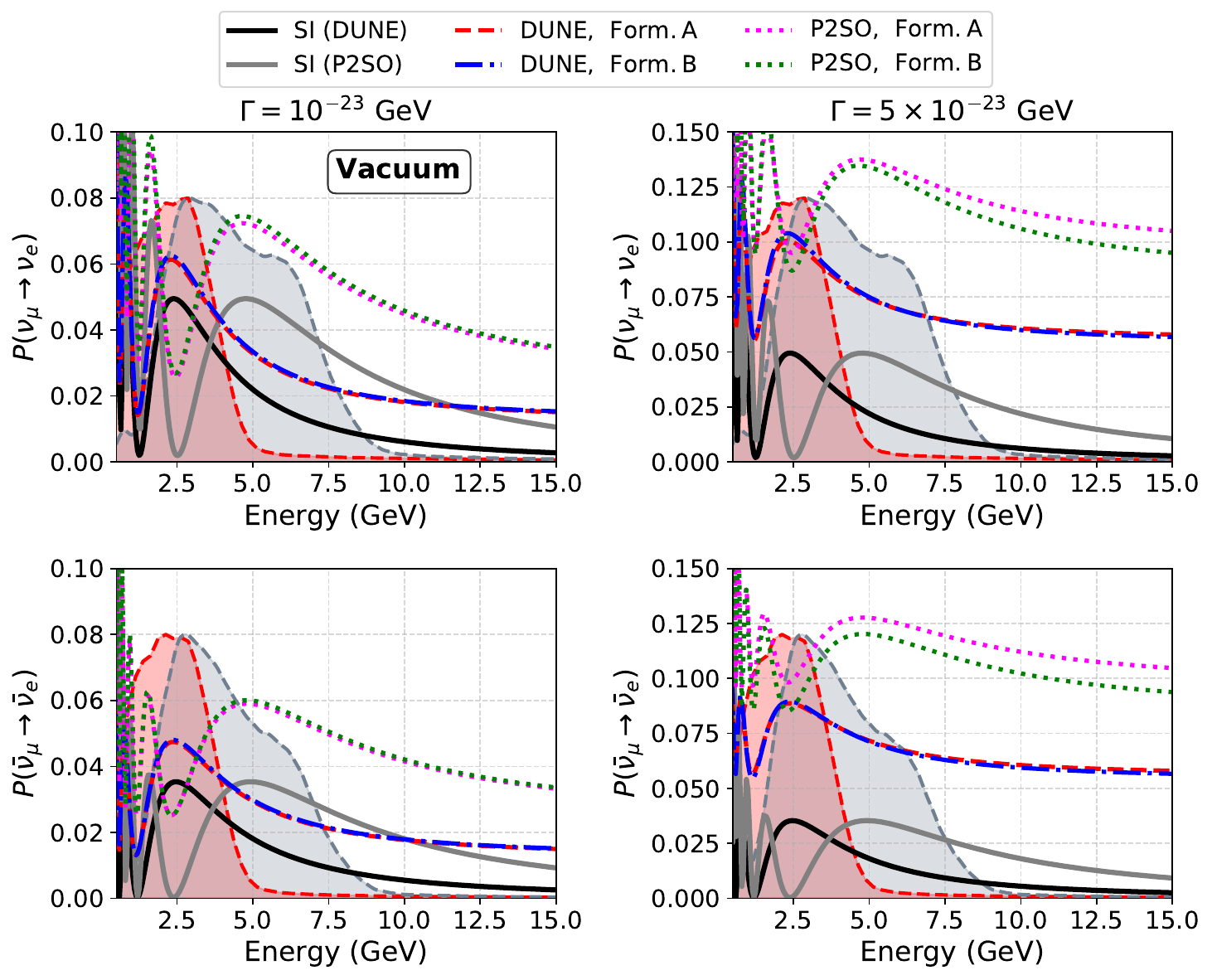}
\caption{Appearance probabilities as a function of energy  for DUNE and P2SO experiments in standard and in presence of the decoherence in vacuum.}
\label{fig:vac-oscillation}
\end{figure}

In Fig. \ref{fig:vac-oscillation}, we have shown the appearance channel probabilities in vacuum as a function of energy for neutrino and antineutrino cases in the upper and lower  rows, respectively. We have considered two benchmark values for the decoherence parameter: $\Gamma = 10^{-23}$ GeV and $5 \times 10^{-23}$ GeV. The plots in the left column are for $\Gamma = 10^{-23}$ GeV and in right column  are for $\Gamma = 5 \times 10^{-23}$ GeV. We have shown the standard interaction (SI) scenario in solid curve and formalism-A and B with dashed curves for both the experiments. The gray (red) shaded region represents the flux of the DUNE (P2SO) experiment. The first thing we notice from this figure is that the probability curves in presence of decoherence are well separated from the standard curves. Further, we note that the probabilities of formalism-A are exactly matching with the probabilities of formalism-B for the value of decoherence parameter  $\Gamma = 10^{-23}$ GeV for both the experiments as well as for both neutrinos and antineutrinos. We have checked that this statement also holds for $\Gamma = 10^{-24}$ GeV. However, when we increase the  value of decoherence parameter to a slightly higher value, e.g., $\Gamma  = 5\times 10^{-23}$ GeV, both formalisms tend to differ from each other. The difference is more for P2SO as compared to DUNE. We have checked that with further increase of the decoherence parameter, deviations between the formalisms are also larger.  From this observation, we can conclude that for small values of $\Gamma$, both the formalisms exactly match in vacuum, however they tend to differ for large $\Gamma$. Next, we  are interested to investigate the situation in presence of matter. 

\begin{figure}
\includegraphics[width=1\linewidth]{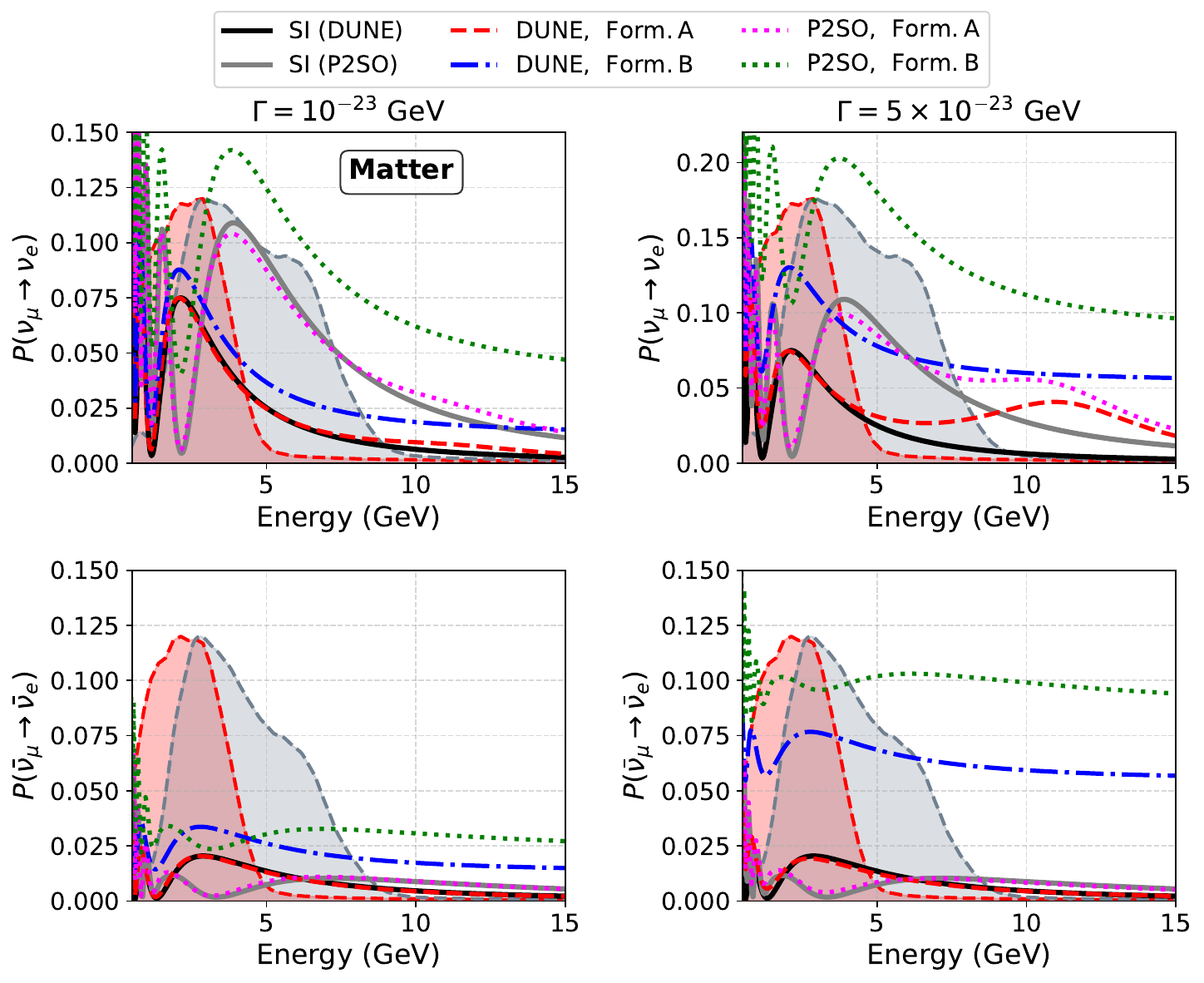}
\caption{Appearance probabilities as a function of neutrino energy  for DUNE and P2SO experiments in standard and in presence of the decoherence in matter.}
\label{fig:mat-osc}
\end{figure}

Figure \ref{fig:mat-osc} depicts the same result as in Fig.~\ref{fig:vac-oscillation}, but for matter. From the figure, we can observe that unlike the vacuum case, even for the choice of $\Gamma = 10^{-23}$ GeV, there is a significant difference between the two formalisms for both the experiments and for both neutrinos and antineutrinos. In fact, for formalism-A, the curves almost coincide with the standard case, and for formalism-B, there is an enhancement in the probability. For a larger value of $\Gamma = 5 \times 10^{-23}$ GeV, the difference between the two formalisms increase further. For this value of $\Gamma$, the purported peak in the probability around 11 GeV appears for the formalism-A as pointed out in Refs.~\cite{Carpio:2017nui}. 

In the following sections, we will study the differences in the sensitivities at the $\chi^2$ level for these two formalisms. In this context, we will also comment if the purported peak in formalism-A contributes to the sensitivity.

\subsection{Sensitivity limits on the decoherence parameters}

\begin{figure}[htbp!]
\includegraphics[width=1\linewidth]{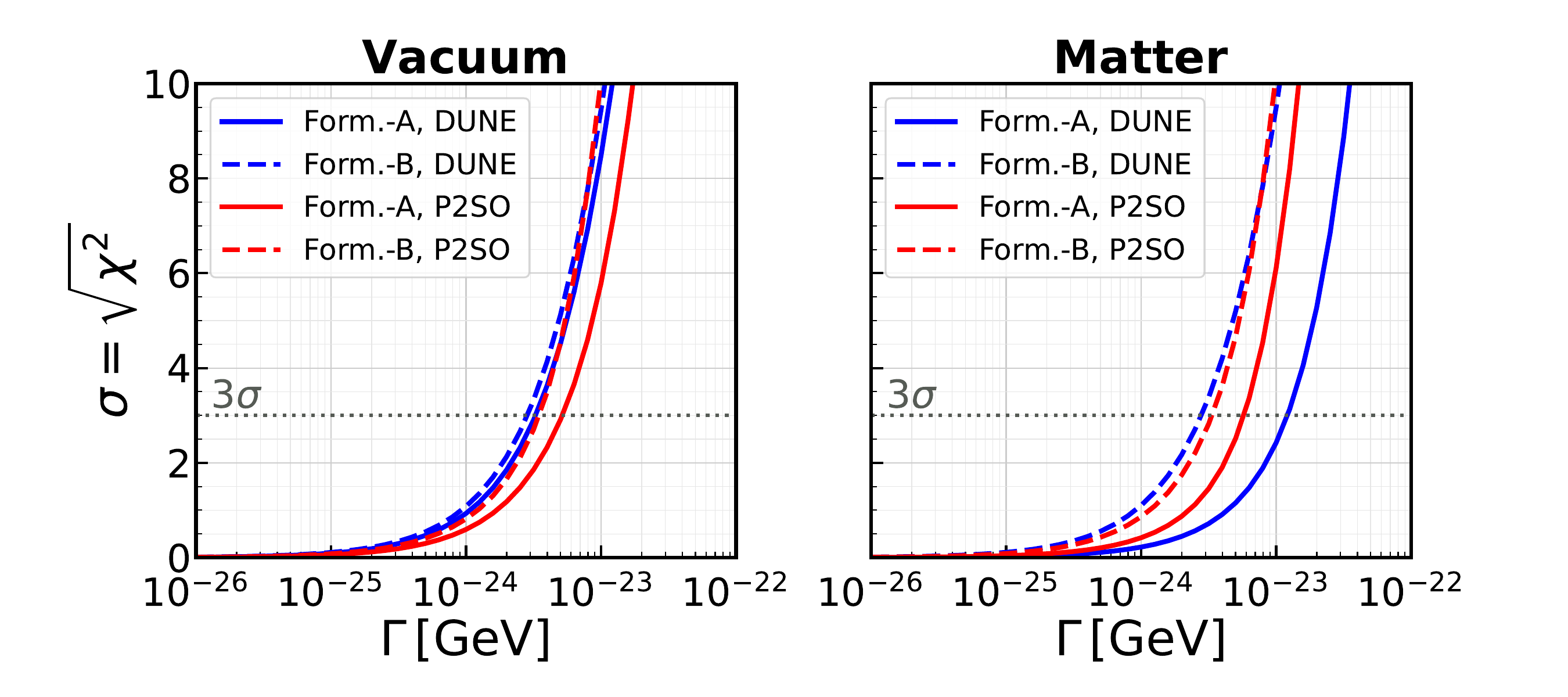}
\caption{Constraints on the decoherence parameters considering formalisms A and B in presence of vacuum and matter for DUNE and P2SO experiments.}
\label{fig:sensitivity-limits}
\end{figure}

In this subsection, we present the sensitivity limits on the decoherence parameters obtained using both formalisms for the DUNE and P2SO experiments. Figure~\ref{fig:sensitivity-limits} shows the sensitivity limits of the decoherence parameter for both the formalisms, considering neutrino propagation in vacuum (left panel) as well as in matter (right panel). The solid and dashed curves correspond to the sensitivities obtained using formalism-A and formalism-B, respectively. The blue (red) curves represent the sensitivity limits obtained from the DUNE (P2SO) experiment. The numerical values of the constraints at $3 \sigma$ C.L. are summarized in Table~\ref{tab:constraints}.

From the figure and the table we see that in vacuum, the constraints for formalism-A and formalism-B are very close in DUNE however, they are a bit different for P2SO. In this case formalism-B provides better constrain on $\Gamma$ for both the experiments. The difference between the formalisms gets significantly enhanced in presence of matter. In this case also formalism-B provides a better bound on the decoherence parameter. Here we would like to emphasize that the purported peak in the probability spectrum which occurs in formalism-A, corresponds to a value of $\Gamma$ which is much larger than the $3\sigma$ bounds on this parameter for both the experiments. Therefore, we can conclude that the purported peak in formalism-A does not contribute in the estimation of the bounds of the decoherence parameter for DUNE and P2SO.

\begin{table}[htbp!]
\centering
\renewcommand{\arraystretch}{1.3}
\begin{tabular}{|c|cc|cc|}
\hline
\multirow{2}{*}{Experiment} 
& \multicolumn{2}{c|}{Formalism--A} 
& \multicolumn{2}{c|}{Formalism--B} \\ \cline{2-5}
& Vacuum & Matter & Vacuum & Matter \\ 
& \multicolumn{2}{c|}{$\Gamma \; [10^{-24}\,\mathrm{GeV}]$} &\multicolumn{2}{c|}{$\Gamma \; [10^{-24}\,\mathrm{GeV}]$} \\ \hline
DUNE  & 3.26 & 12.05  & 2.83 & 2.77 \\ \hline
P2SO  & 5.14 & 5.72  & 3.45 & 3.33 \\ \hline
\end{tabular}
\caption{Constraints on the decoherence parameter $\Gamma$ from DUNE and P2SO for vacuum and matter cases under formalism--A and formalism--B.}
\label{tab:constraints}
\end{table}

\subsection{Sensitivity to mass ordering, octant and CPV}

\begin{figure}
\includegraphics[scale=0.22]{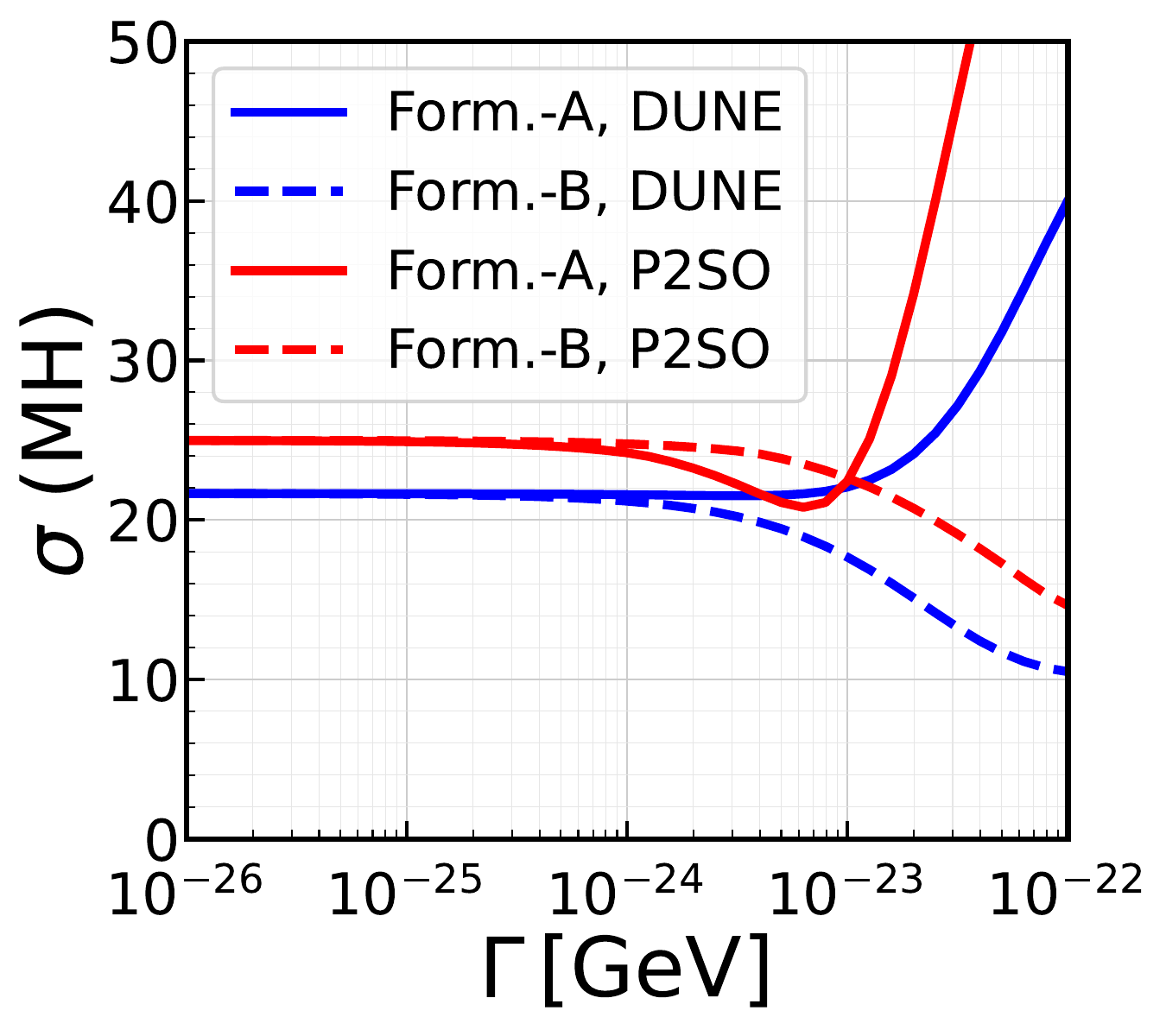}
\includegraphics[scale=0.22]{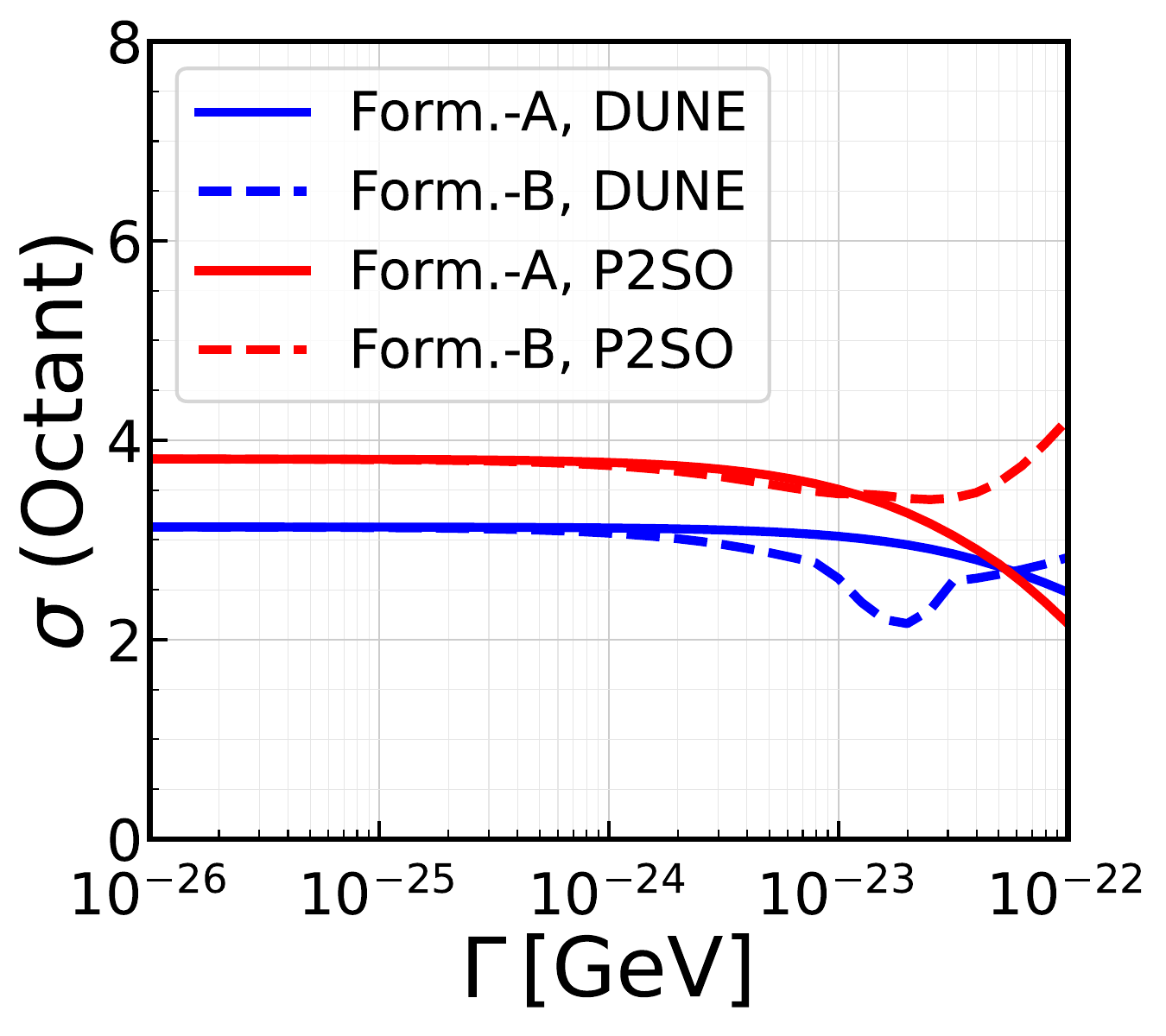}
\includegraphics[scale=0.22]{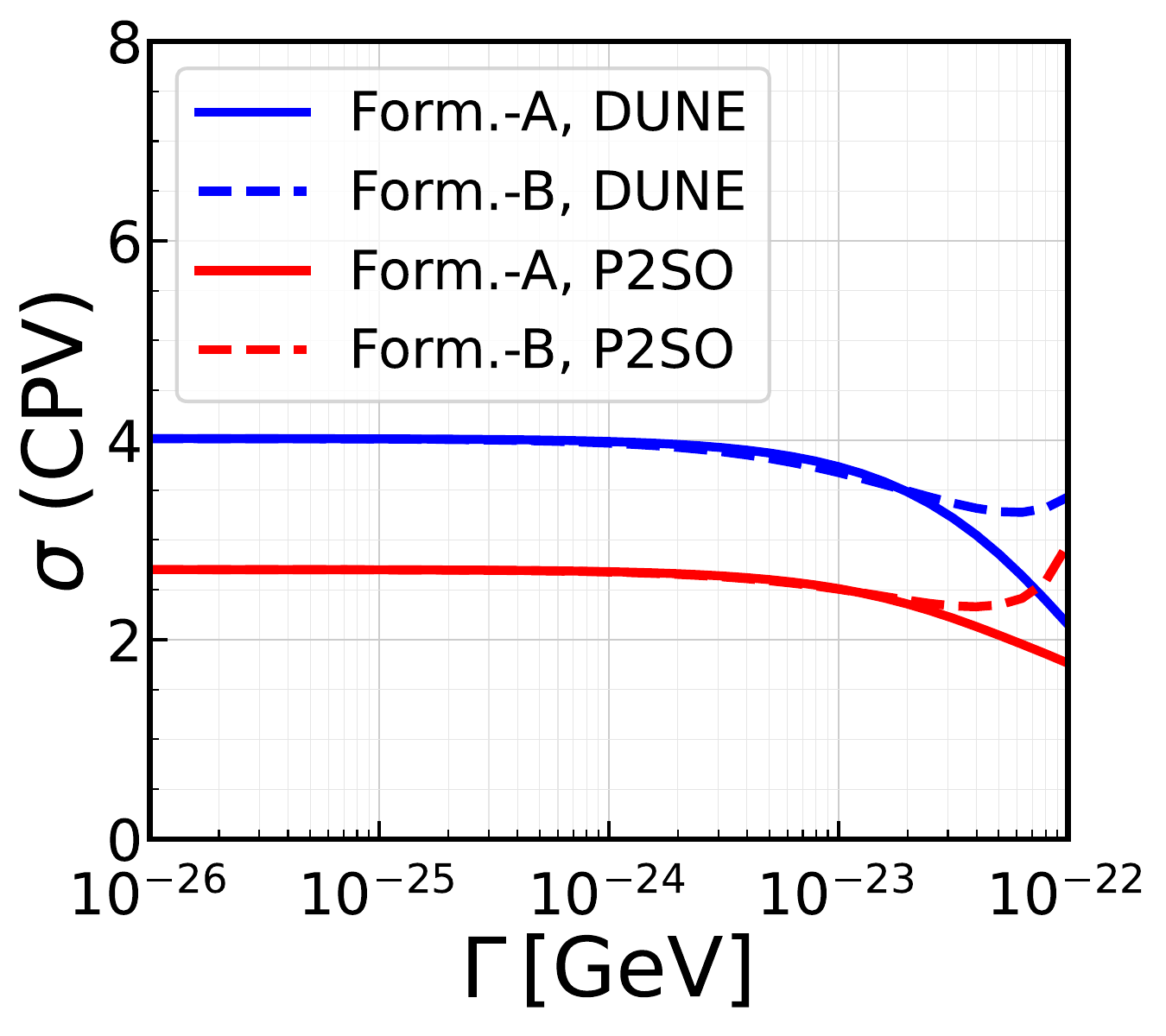}
\caption{Dependence of the mass ordering, octant, and CPV sensitivities on the decoherence parameter in formalism-A (solid curves) and in formalism-B (dashed curves). The left, middle, and right panels show the mass ordering, octant, and CPV sensitivities, respectively. Results for DUNE and P2SO are shown in blue and red, respectively.}
\label{sens}
\end{figure}

In this section, we will show the impact of different formalisms in determining the sensitivity to neutrino mass ordering, octant of $\theta_{23}$ and CP violation (CPV) in presence of decoherence. We have presented these results in Fig.~\ref{sens} as a function $\Gamma$. The left/middle/right panels shows the sensitivity for mass ordering/octant/CPV in terms of $\sigma = \sqrt{\chi^2}$. The mass ordering sensitivity quantifies how effectively an experiment can distinguish the correct neutrino mass ordering from the wrong mass hierarchy whereas the octant sensitivity quantifies the ability of an experiment to resolve the octant degeneracy. Finally the CP violation sensitivity quantifies the ability of an experiment to distinguish CP-violating effects from CP-conserving scenarios, corresponding to $\delta_{\rm CP} = 0$ or $\pm \pi$. For the mass ordering we have considered normal as true ordering whereas for octant we have assumed lower octant to be the correct octant. For CPV, we have chosen $\delta_{\rm CP} = 212^\circ$ as the true value. In each panel the solid curves correspond to formalism-A whereas the dashed curves correspond to formalism-B. Blue curves are for DUNE whereas red curves are for P2SO. 

From the left panel, we see that for small values of the decoherence parameter, up to $\Gamma \lesssim 10^{-24}\,\mathrm{GeV}$, the mass ordering sensitivities obtained using the two formalisms are nearly identical and they are close to the standard oscillation values. As the value of the decoherence parameter increases further, the sensitivity obtained in formalism-A increases compared to the standard scenario, whereas the sensitivity in formalism-B gradually decreases relative to the standard case for both DUNE and P2SO experiments. However, a further increase in the decoherence parameter within formalism-B may lead to an enhancement in sensitivity. It is important to note that such large values of the decoherence parameter lie beyond the current $3\sigma$ sensitivity limits. Moreover, the difference between the two formalisms is more pronounced for P2SO than for DUNE at higher values of the decoherence parameter above $10^{-23}~\mathrm{ GeV}$.

In the middle panel, the octant sensitivities obtained from both formalisms are identical for lower values of the decoherence parameter up to $\sim 10^{-24}~\mathrm{GeV}$. Formalism-B exhibits an interesting behavior: the sensitivity initially decreases with increasing $\Gamma$ and then rises for higher values. In contrast, formalism-A shows a gradual and monotonic decrease in sensitivity with increasing $\Gamma$. For larger values of the decoherence parameter, the sensitivity in formalism-B becomes higher than that in formalism-A. Furthermore, the difference between the two formalisms is more pronounced for P2SO compared to DUNE.

In the right panel, similar to the mass hierarchy and octant sensitivities, the CPV sensitivity remains nearly unaffected by decoherence effects for $\Gamma \lesssim 10^{-24}~\mathrm{GeV}$. Beyond this range, the CPV sensitivity decreases relative to the standard scenario for both formalisms. However, in formalism-B, the sensitivity begins to increase again at higher values of $\Gamma \sim 2 \times 10^{-23}~\mathrm{GeV}$. A clear separation between the CPV sensitivities of the two formalisms emerges for $\Gamma \gtrsim 2 \times 10^{-23}\mathrm{ ~GeV}$.

\section{Summary}
\label{summary}

In this work we have investigated the impact of  different formalisms of quantum decoherence in determining the sensitivities of the two future long-baseline experiments DUNE and P2SO. In formalism-A, the decoherence matrix is defined in a matter mass eigenstate basis as diagonal
and energy independent whereas in formalism-B, the decoherence matrix is defined in the vacuum mass eigenstate basis and one requires to rotate it to matter mass basis via an unitary transformation. 

First we show that in vacuum, for $\Gamma \lesssim 10^{-23}\,\mathrm{GeV}$, both the formalisms yield identical probabilities and for larger values of $\Gamma$, deviations between the two formalisms become apparent. In presence of matter, even at $\Gamma \lesssim 10^{-23}\,\mathrm{GeV}$, both the formalisms yield different probabilities. In formalism-A, we observed a peak around 11 GeV for  $\Gamma = 5 \times 10^{-23}\,\mathrm{GeV}$ which is absent in formalism-B. While studying the impact on determining the upper bound on the decoherence parameter, we found out that in vacuum, the constraints for formalism-A and formalism-B are very close in DUNE however, they are a bit different for P2SO. The difference between the formalisms get significantly enhanced in presence of matter. Finally, we showed that  both the formalisms yield very different sensitivities for the determination of neutrino mass ordering, octant of $\theta_{23}$ and CP violation in presence of decoherence.  For mass ordering, formalism-A provides better sensitivity than formalism-B, whereas for octant and CP violation, formalism-B provides better sensitivity.

In summary, we would like to point out that while determining sensitivities of long-baseline experiments in presence of quantum decoherence, it is very important to choose the correct formalism, specially in the case where the matter effect is significant. As formalism-B, takes into account the matter effect properly, we believe the sensitivities obtained with this formalism are more realistic. We further reiterate that this work represents the views of the authors and not the DUNE Collaboration.

\section*{Acknowledgments}
RM (Rudra Majhi) would like to acknowledge Odisha State Higher Education Council, Govt. of Odisha for the support under Mukhyamantri Research and Innovation (MRIP)-2024 (24EM/PH/102). KP acknowledges the University Grants Commission(UGC) for supporting his doctoral work. The work of MG has been in part funded by Ministry of Science and Education of Republic of Croatia Grant No. PK.1.1.10.0002, Swiss National Science Foundation (SNSF) and Croatian Science Foundation (HRZZ) under grant MAPS IZ11Z0$\_$230193 and European Union under the NextGenerationEU Programme. Views and opinions expressed are however those of the author(s) only and do not necessarily reflect those of the European Union. Neither the European Union nor the granting authority can be held responsible for them.  

\bibliography{biblio.bib}
\end{document}